\documentclass[11pt,a4paper]{article}
\usepackage[left=20mm, top=20mm, right=20mm, bottom=20mm, nohead=5mm, nofoot=5mm]{geometry}
\usepackage[utf8]{inputenc}
\usepackage{authblk}
\usepackage{indentfirst}
\usepackage{misccorr}
\usepackage{graphicx}
\usepackage{amsmath}
\usepackage{amssymb}
\usepackage{multicol}
\usepackage{bm}
\usepackage{longtable}
\usepackage{pdflscape}
\usepackage{epstopdf}
\usepackage{multirow}
\usepackage{adjustbox}
\usepackage{amsfonts}
\usepackage{fancyhdr}
\usepackage{ifthen}
\usepackage{fancyheadings}
\usepackage{setspace}
\usepackage{xcolor}
\usepackage[round,authoryear,comma,sort&compress,numbers]{natbib}
\usepackage[toc,title,page]{appendix}
\usepackage[hidelinks]{hyperref}
\usepackage{url}

\hypersetup{
    colorlinks=true,
    linkcolor=green,
    citecolor=blue,
    filecolor=magenta,
    urlcolor=cyan,
    pdftitle={Sharelatex Example},
    bookmarks=true,
    pdfpagemode=FullScreen,
}

\newcommand{\aap}{Astronom. and Astrophys.}

\newcommand{\apj}{Astrophys. J.}
\newcommand{\apjs}{Astrophys. J. Suppl.}

\newcommand{\araa}{Annu. Rev. Astronom. Astrophys.}

\newcommand{\mnras}{Monthly Notices Roy. Astronom. Soc.}

\newcommand{\pasp}{Publ. Astronom. Soc. Pacific}

\fancypagestyle{plain}{\chead{\texttt{\textcolor{brown}{Communications of BAO, Vol. ?, Issue ?, 20??, pp. ???-???}}}}
\title{\textbf{The Origin of LAMOST J1109+7459}}
\author[1]{Nour Aldein Almusleh}
\author[1]{Yazan Khrais}
\affil[1]{\scriptsize Department of physics al Al-Bayt university , al mafraq , 130040 Jordan }

\author[2]{Ali Taani  \thanks{ali.taani@bau.edu.jo, Corresponding author}}
\affil[2]{\scriptsize Physics Department, Faculty of Science, Al-Balqa Applied University,Jordan}

\begin{document}
\pagestyle{empty}
\newpage
\pagestyle{fancy}
\label{firstpage}
\date{}
\maketitle

\begin{abstract}
We report a comprehensive Chemo-dynamical analysis of LAMOST J1109+0754, a relatively bright (V = 12.8), extremely metal-poor ([Fe/H] = $-3.17$), and prograde ($J_\phi$ and $V_\phi$ $> 0$) star, with a strong \textit{r}-process enhancement ([Eu/Fe] = $+$0.94 $\pm$ 0.12, [Ba/Fe] = $-$0.52 $\pm$ 0.15). 31 chemical abundances (from Lithium to Thorium) were derived. We suggest a possible progenitor with stellar mass of 13.4-29.5 M$_\odot$. We argue that J1109+0754 is representative of the main \textit{r}-process component due to the well agreement with the scaled-solar \textit{r}-process component. We analyze the orbital history of this star in a {\it time-varying Galactic potential}, based on a Milky-Way analogue model extracted from \texttt{Illustris-TNG} simulations. Using this model, we carry out a statistical estimation of the phase-space coordinates of J1109+0754 at a young cosmic age. Collectively, the calculated motions, the derived chemistry, and the results from the cosmological simulations suggest that LAMOST J1109+0754 most likely formed in a low-mass dwarf galaxy, and belongs to the Galactic outer-halo population.
\end{abstract}

\section{Introduction}

Our universe is made up of a substantial number of elements (from hydrogen to oganesson) and their isotopes. This fact inspired astronomers and nuclear physicists, for more than six decades, to investigate the major physical processes and conditions that led to produce this wide variety of chemical species. Since the pioneer studies of \citet{1957RvMP...29..547B} and \citet{1957PASP...69..201C}, there has been a large number of studies in the literature focusing on the astrophysical site(s) of the rapid neutron capture (\textit{r})-process. However, the problem has not been solved, but some promising mechanisms were proposed: (i) the innermost ejecta of regular core-collapse supernovae \citep[e.g.,][]{2010ApJ...712.1359F}, (ii) outer layers of supernova explosions \citep[e.g.,][]{2014JPhG...41d4002Q}, (iii) magneto-rotational jet-driven supernovae \citep[e.g.,][]{2018JPhG...45h4001O}, and (iv) neutron stars (NSs) mergers \citep[e.g.,][]{2017ARNPS..67..253T}. 


The science of stellar archaeology is built on two fundamental assumptions: (i) Population II stars preserve in their atmosphere the chemical composition of the individual or a few Supernovae (SN) yields of the previous Population III \citep[the so-called Population III chemical fingerprint,][]{Mardini_2019b} and (ii) Accurate chemical abundances (from lithium to uranium) can be acquired from high-resolution spectra. Therefore, metal-poor stars can contribute to reveal the first nucleosynthesis enrichment in the universe. In particular, the spectroscopic analysis of individual galactic halo stars with enhancement in \textit{r}-process elements and lack/low \textit{s}-process elements\footnote{The criterion [Ba/Fe] $>$ 0 is included into the definitions of \textit{r}-process-enhanced stars to avoid any possible contribution from the \textit{s}-process.} (the so-called \textit{r}-process-enhanced stars) bring us closer to solve the long-standing \textit{r}-process puzzle \citep[for a selected list see e.g.,][and references therein]{2020ApJ...897...78P}. 

For the aforementioned reasons, increasing the numbers of the known \textit{r}-process-enhanced metal-poor stars (i.e., expanding the chemical inventory of these stellar objects) has been one of the major interesting research subjects for individual researchers \citep[e.g.,][]{Mardini_2019a} and large projects such as the Hamburg/ESO and the \textit{R}-Process Alliance. In addition, \citet{2005ARA&A..43..531B} have proposed that \textit{r}-process-enhanced stars can be classified, using the observed [Eu/Fe] ratio, into two main distinctive populations (i) \textit{r}-I stars  with $0.3 \leqslant $ [Eu/Fe] $\leqslant 1.0$ and (ii)  \textit{r}-II with [Eu/Fe] $>1.0$. Based on the database for metal-poor stars \citep[JINAbase,][]{2018ApJS..238...36A}\footnote{https://jinabase.pythonanywhere.com} , the current statistics of \textit{r}-I and \textit{r}-II are $\sim$30 and 125, respectively. 


Besides the chemical compositions of individual galactic stars, investigating the positions and kinematics of these stars will advance our understanding of the formation and evolution of our galaxy. During the past 30 years much more information has become available on the structure of the stellar halo of the Milky Way. These efforts suggested that the galactic halo is made-up of two distinguished populations of stars; the inner and outer stellar halo \citep[e.g.,][]{1990MNRAS.242...10S,2019AstBu..74..464M,2019AN....340..847T,2019JPhCS1258a2029T,2019JPhCS1258a2024M,2019RAA....19...12T,2020arXiv200203011T}. Furthermore, in the era of the second data release of the Gaia mission \citep[Gaia DR2;][]{2018A&A...616A...1G}, several studies are currently being expended investigating the kinematics and chemical abundance patterns of  \textit{r}-process-enhanced stars. These efforts attempt to combine the chemical abundance patterns, observed in \textit{r}-process-enhanced stars, with kinematics derived from Gaia DR2 observations and results from cosmological simulations to assess the formation mechanisms and determine the role of  \textit{r}-process-enhanced stars as tracers of the accretion history of the galactic halo, i.e. the time at which these stars were ejected into our Galaxy.

\section{Data, Results, and Discussions} \label{sec:data}

\begin{figure}[ht]
\centering
  \includegraphics [width=2.8in] {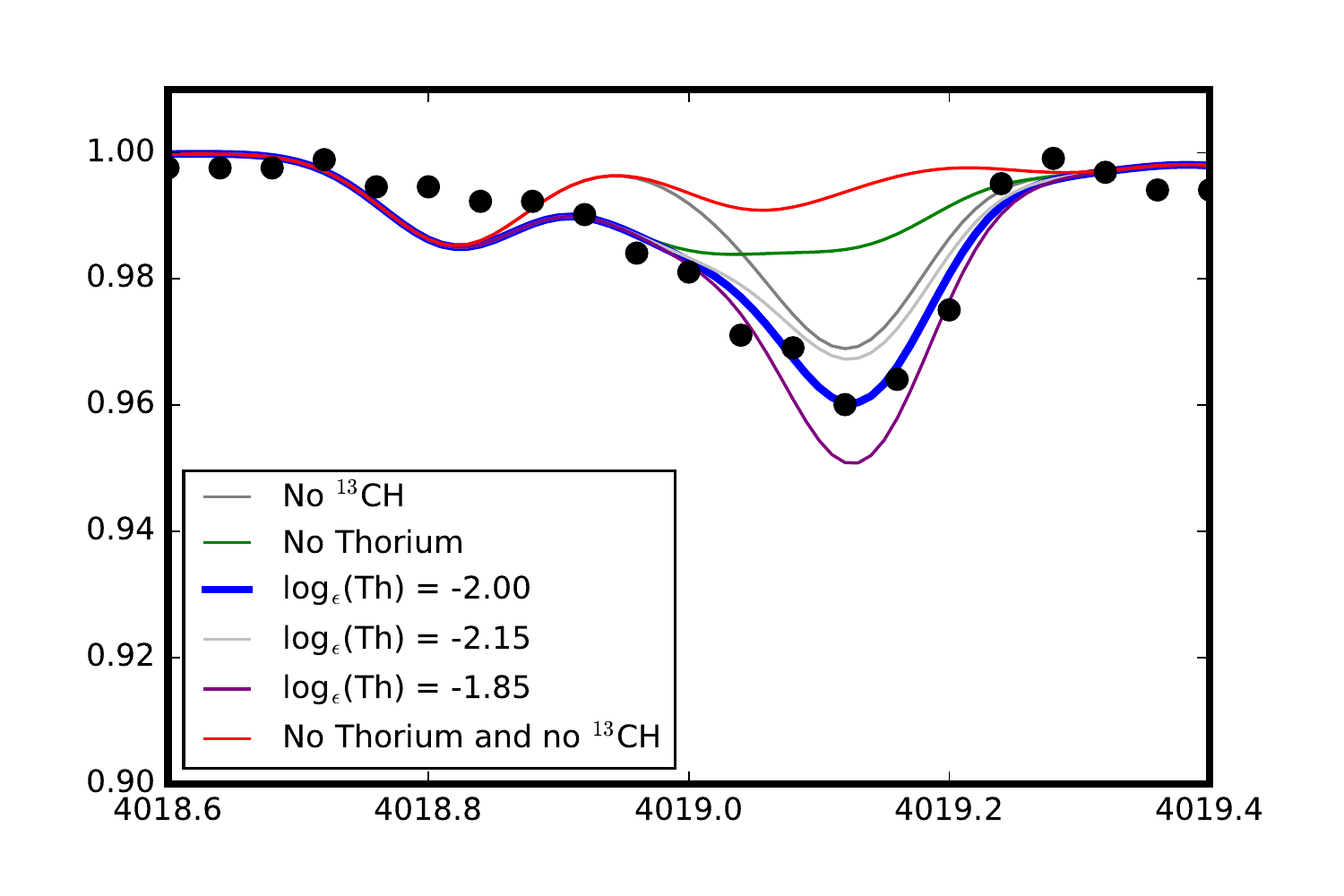}
  \includegraphics [width=2.8in,height=1.8in]{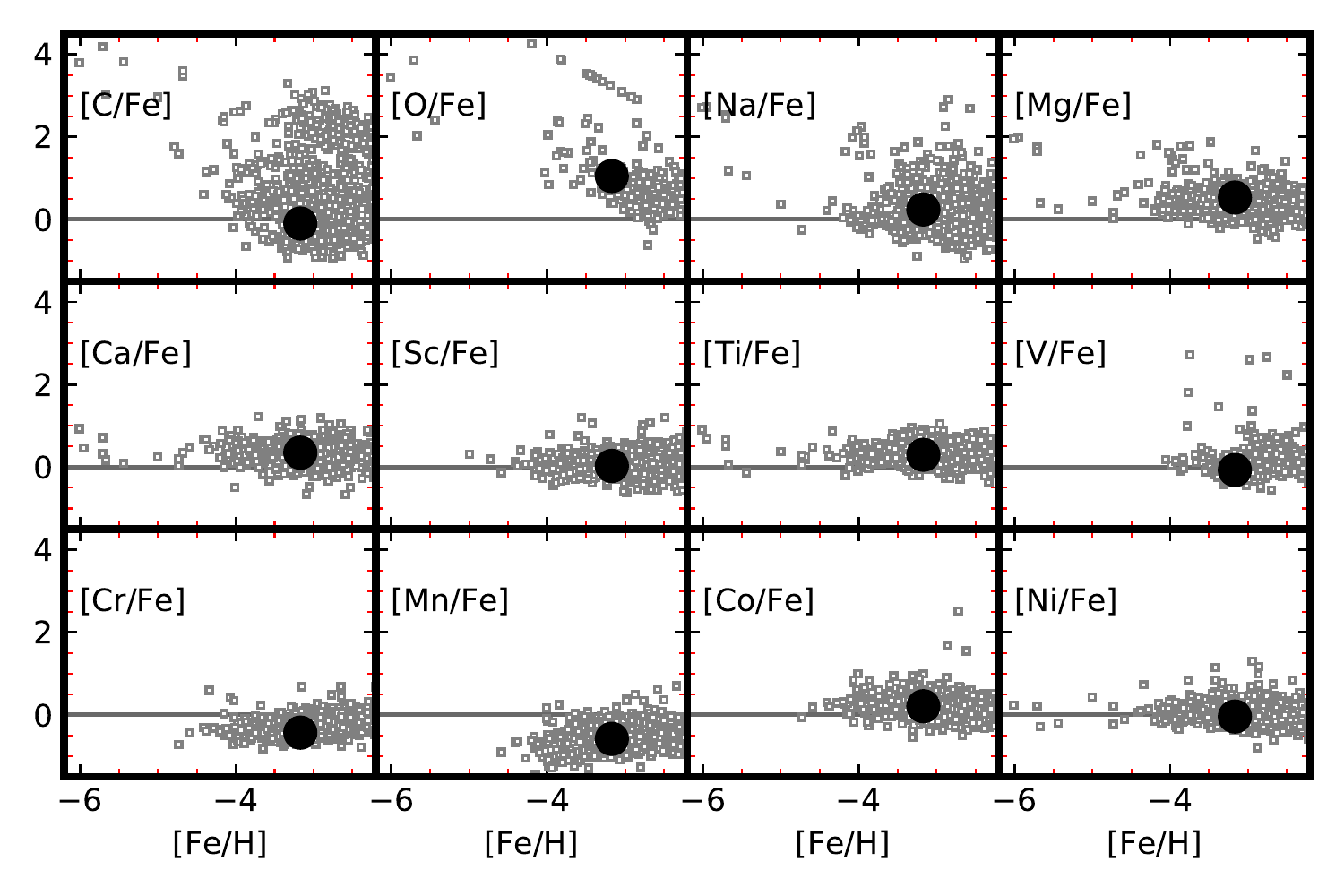}
  \includegraphics [width=2.8in] {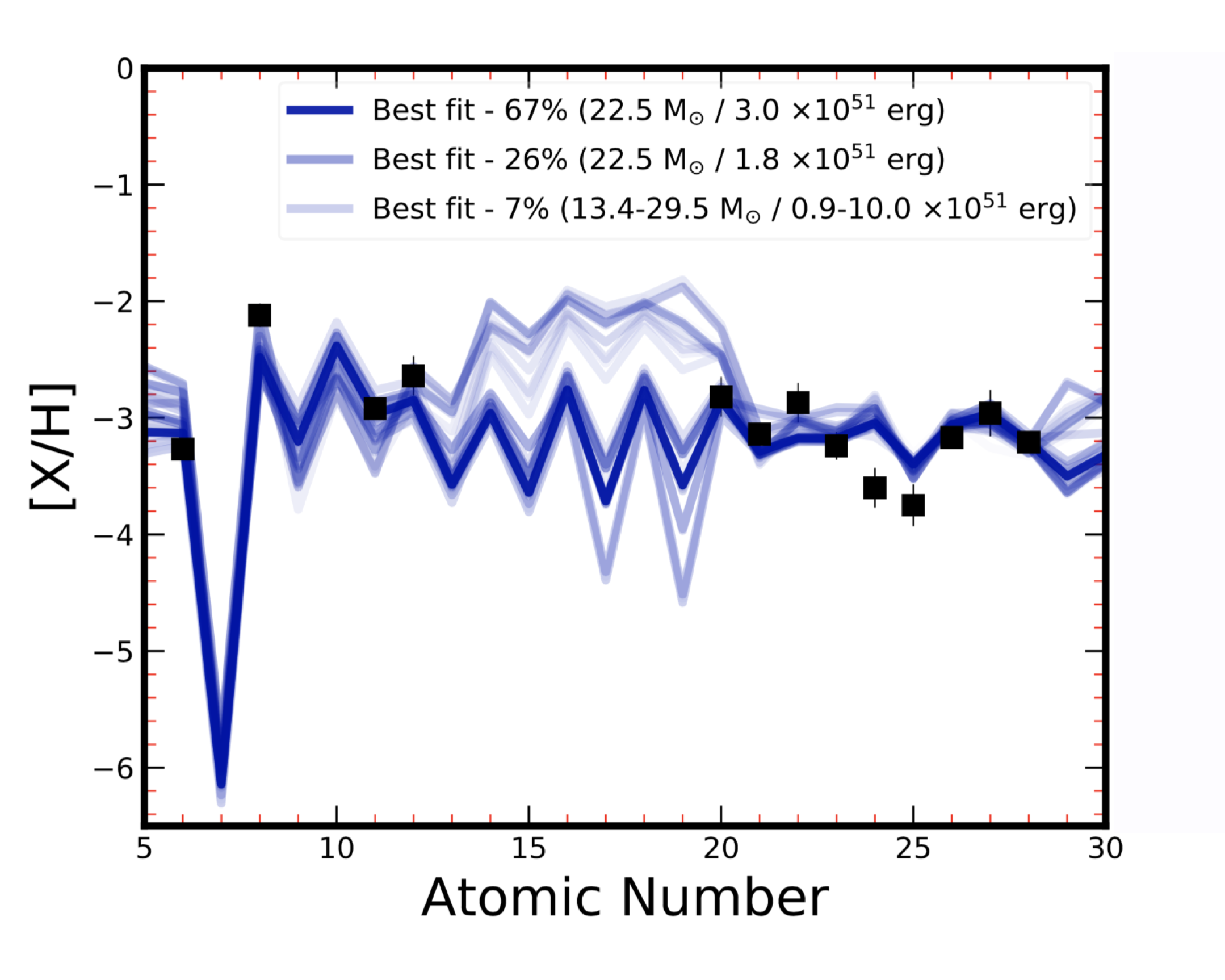}
  \includegraphics [width=2.8in,height=2.1in, trim = 1cm 0.0cm 0.1cm 0.1cm] {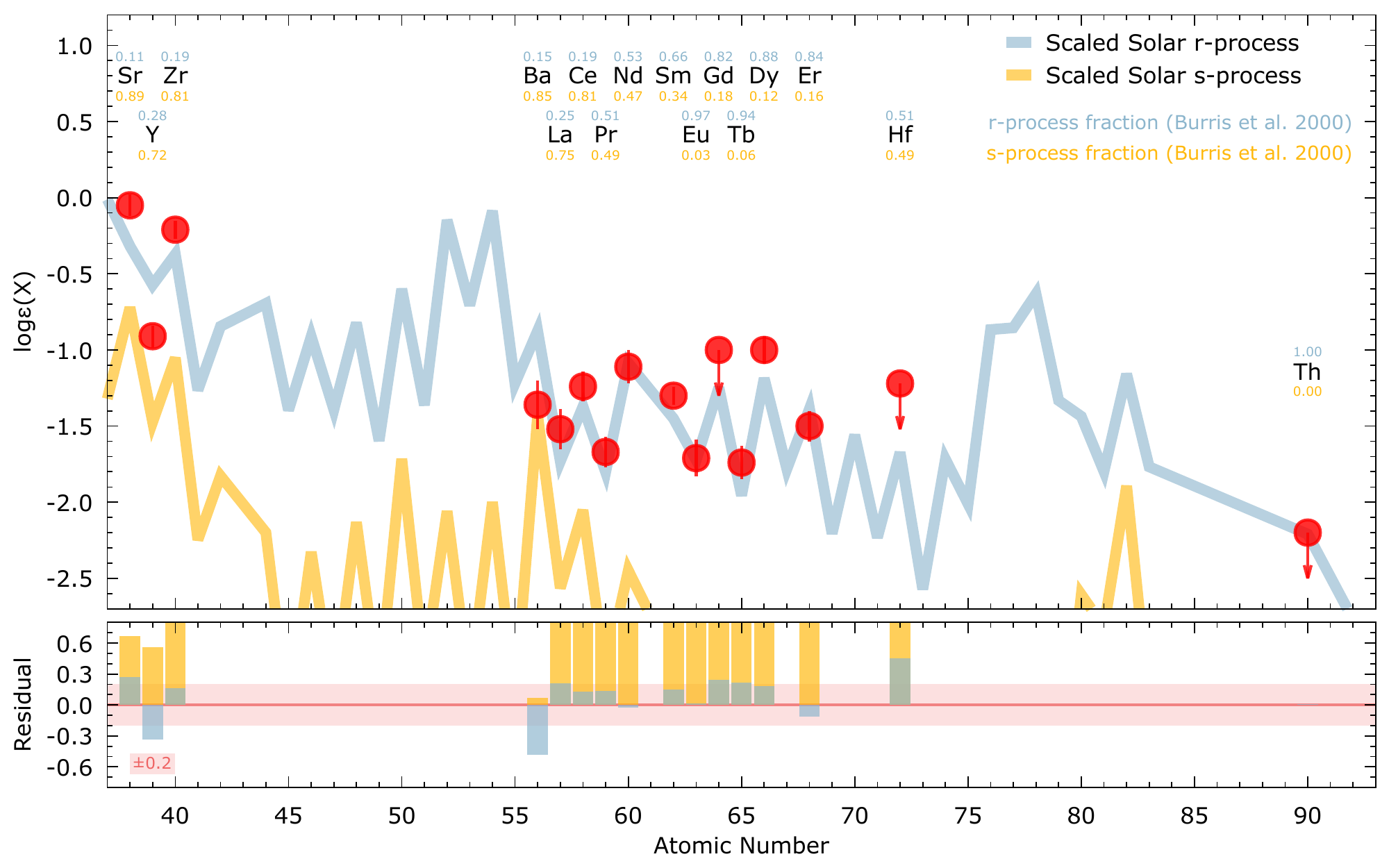}
  \caption{Upper left panel: portion J1109+0754 spectrum was used to derive the thorium (Z = 90) abundance and the associated uncertainty. Upper right panel: observed [X/Fe] ratios for elements up to the iron peak, as a function of [Fe/H]. Lower left panel: the observed [X/H] abundance ratios of J1109+0754 (filled black squares), as a function of atomic number, overlaid with the matched predicted nucleosynthetic SNe models. Lower right panel: heavy-element abundance pattern of J1109+0754 (filled circles), overlaid with the scaled Solar System abundances (SSSA).} \label{fig:one}
\end{figure}

In 2015, we observed J1109+0754 using Automated Planet Finder (APF) Telescope. We used IRAF to carry out a standard echelle data reduction. We measured the Radial Velocity using the same method presented in \citet{Mardini_2019b}, by cross-correlating the high-resolution reduced spectrum against synthesized template of the same spectral type, making use of the Mg I triplet at 5160–5190,\AA. We used MOOG to determine stellar parameters and chemical abundances for 31 elements including Thorium.

The upper left panel of Figure \ref{fig:one} shows an illustrative example of spectral synthesis approach. The portion J1109+0754 spectrum was used to estimate the thorium abundance and its associated uncertainty. The upper right panel of Figure \ref{fig:one} shows the distribution of these abundances (large solid points), as you can see the light elements abundances of J1109+0754 do not scatter from the general trend observed in other literature metal-poor field stars (small opened gray squares). The lower left panel of Figure \ref{fig:one} speculate the stellar mass and the SN explosion energy of the progenitor of J1109+0754. The fitting result of this exercise supports the conclusion presented in \citet{Mardini_2019b}, which suggests that a stellar mass $\sim$ 20 M$_{\odot}$ may reflect the initial mass function of the first stars. Furthermore, it brings the possibility as to whether more massive SN might be more energetic and therefore destroy their host halo and not allow for EMP star formation afterward. The lower right panel of Figure \ref{fig:one} shows the heavy elements patterns overlaid with the scaled Solar components. The observed deviations of Sr, Y, and Zr from the  scaled r-process peak indicate that the production of these elements (first r-process peak) is likely to be separated from the second and third r-process peaks. The universality of the r-process is shown in J1109+0754 due to the good agreement between the abundances of the elements with Z $> 56$ the scaled Solar r-process component. This suggest that there is no clear contribution from the s-process. Therefore, we argue  that J1109+0754 can be used as a representative of the  main component of the r -process \citep{2020arXiv200912142M, 2020ApJ...902..125A}.

\begin{figure}[ht]
\centering
  \includegraphics [width=2.8in] {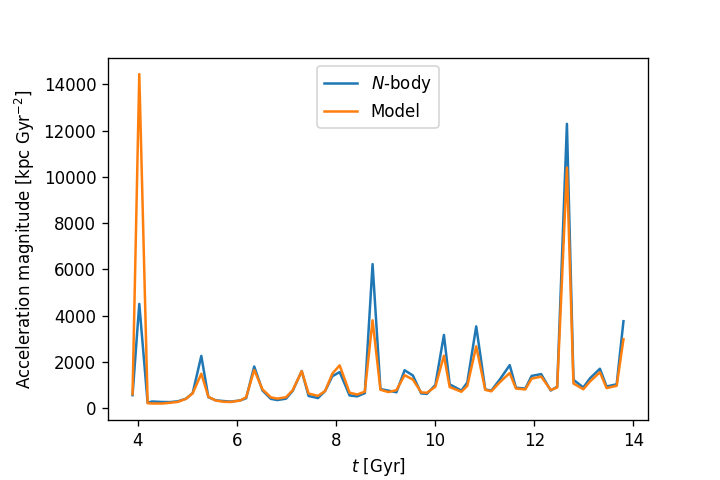}
  \includegraphics [width=2.8in] {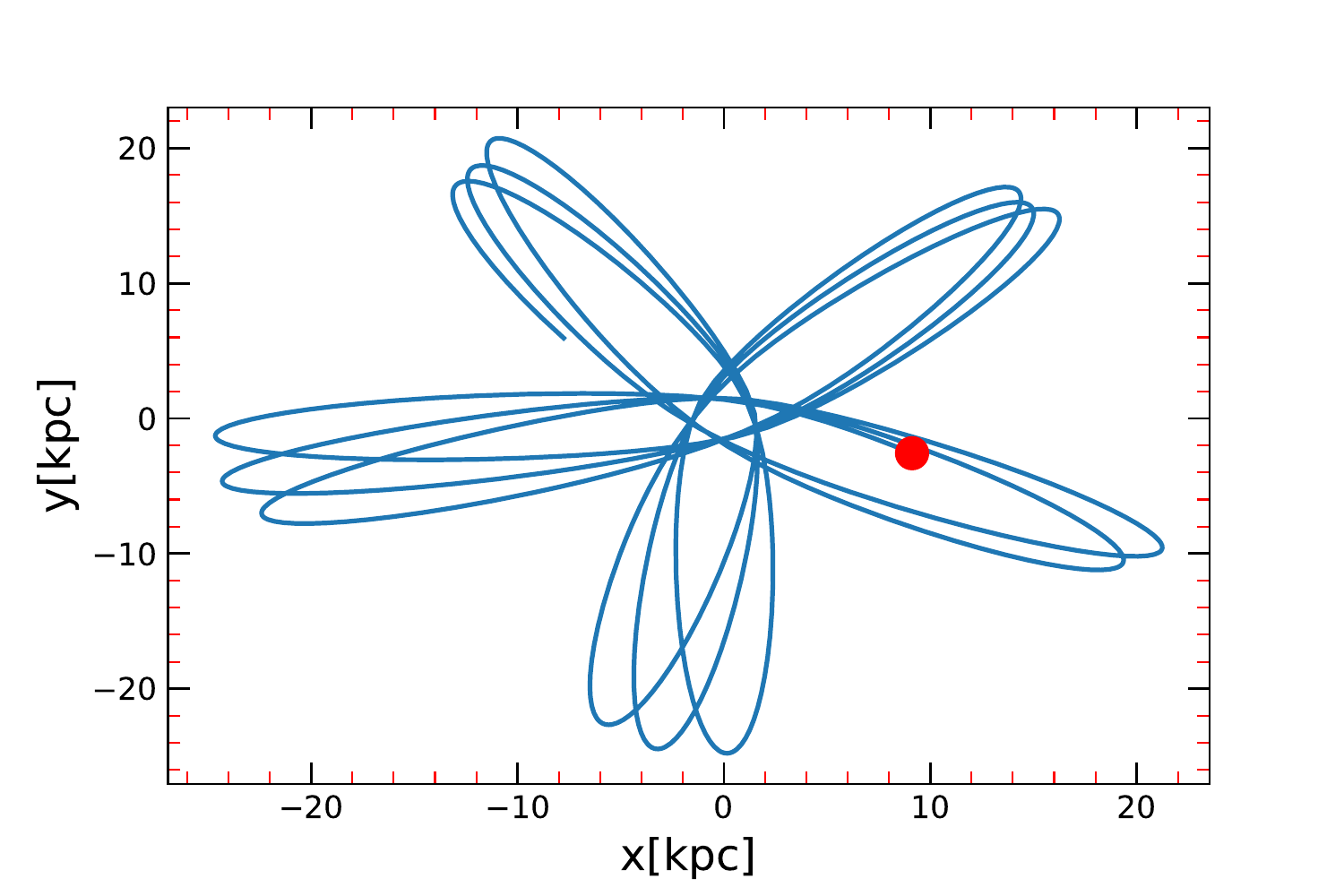}
  \caption{Left panel: the gravitational acceleration of J1109+0754 along its orbit using our model as well as the actual particle distribution from the Illustris-TNG simulation. Right panel: the last 10 periods orbit of J1109+0754, in x-y projection integrated in time for 4.2 Gyr}  \label{fig:two}
\end{figure}

The advent of the Gaia mission has fundamentally changed our view of the nature of the Milky Way. However, unraveling the full kinematic signature of J1109+0754 requires a less-idealized and more-realistic time-dependent galactic gravitational potential. We use the Illustris-TNG TNG100 simulation box, characterized by a length of $\sim$110\,Mpc to select potential Milky Way analogs. We then pared down an initial list of over 2000 candidates into using some sophisticated stellar parameters cuts. The best-fitting parameters are found for each snapshot using a smoothing spline fitting procedure. To obtain the gravitational force as a function of time, we interpolate between each snapshot’s parameters. In order to show that our NFW + Miyamoto-Nagai model provides a reasonable description of the time evolving Galactic analogues, we calculated the gravitational acceleration of the star along its orbit using our model as well as the actual particle distribution from the Illustris-TNG simulation. This shows (see Figure \ref{fig:two}) at the very least that our orbital integration of this particular star is approximately correct. To produce the figure, we used the orbital data from a randomly chosen integration in a randomly chosen MW analogue, and interpolated the spatial position at times where the full snapshots from the cosmological simulation were available. Then, the gravitational force was calculated at those spatial points using both our model routines, and direct summation. As one can see, the two calculations match quite closely, with relatively large error occurring only at particular points where the particle is closest to the centre of the halo (where naturally complex substructure is not fully captured by the simple model). This is especially the case in the second snapshot at around 4\,Gyr; at this point in time the system has not yet settled into a simple disk+halo (note also that our integration is backward, so this datapoint has no practical importance in our orbit calculations). Barring this data point, the mean relative deviation between the model and the simulation is 0.04, and the standard deviation is 0.13. We generated 10,000 sets of the six-dimensional phase space coordinates from the corresponding measurement errors. This exercise shows that J1109+0754 has bounded (E= $-87.05$ $\times$$10^{3}$ km$^{2}$ s$^{-2}$), non-planar (Z$_{max}$=10.87\,kpc), and eccentric orbit (e= 0.84). Moreover, J$_\phi$ and V$_\phi$ values (683.68\,kpc km s$^{-1}$ and 73.72\,km s$^{-1}$, respectively) suggests that J1109+0754 is a  prograde star. The right panel Figure \ref{fig:two} shows the last 10 periods orbit of J1109+0754, in x-y projection integrated in time for 4.2 Gyr.

\section{Conclusions} \label{concs}

In this study, we carry out a statistical estimation of the stellar masses of the progenitors, the phase space coordinates, and orbital backward-time integrations of the relatively bright (V = 12.8), extremely metal-poor ([Fe/H] = $-3.17$), and prograde ($J_\phi$ and $V_\phi$ $> 0$) star, with a strong \textit{r}-process enhancement ([Eu/Fe] = $+$0.94 $\pm$ 0.12, [Ba/Fe] = $-$0.52 $\pm$ 0.15) LAMOST J1109+0754. The direct comparison of the observed atmospheric chemical abundances and the predicted SN yields suggest possible progenitors with stellar mass span the range of 13-22\,M$_{\odot}$. In addition, the calculated motions suggest that our star is an outer-halo member. Furthermore, its peculiar atmospheric chemical composition suggest that our stars are might belong to a low-mass dwarf galaxy and have been accreted at a young cosmic age. The action-space map of our sample suggest that this \textit{r}-process-enhanced star does not match the numerical criteria of $Gaia$-Sausage and $Gaia$-Sequoia remnant stars, but, it suggests that another accretion event might be responsible for the contribution of these \textit{r}-process-enhanced stars to the Milky-Way.

To investigate the accretion scenario of \textit{r}-process-enhanced stars, we strongly recommend future identification and observations to increase the numbers of these peculiar stars so that their full kinematics can be investigated in more detail, and thus be used to improve our understanding of their origin.

\section*{\small Acknowledgements}
\scriptsize{This work was supported by the National Natural Science Foundation of China under grant Nos. 11988101 and 11890694, and National Key R\&D Program of China No.2019YFA0405502. This work has made use of data from the European Space Agency (ESA) mission {\it Gaia} (\url{https://www.cosmos.esa.int/gaia}), processed by the {\it Gaia} Data Processing and Analysis Consortium (DPAC, \url{https://www.cosmos.esa.int/web/gaia/dpac/consortium}). Funding for the DPAC has been provided by national institutions, in particular the institutions participating in the {\it Gaia} Multilateral Agreement.

\bibliographystyle{ComBAO}
\nocite{*}
\bibliography{J1109}

\end{document}